# Computational Simulation and 3D Virtual Reality Engineering Tools for Dynamical Modeling and Visualization of Composite Nanomaterials

L.V. Bochkareva[1], M.V. Kireitseu[1,4], G.R. Tomlinson[4], H. Altenbach[4], V. Kompis[2], and D. Hui[5]

[1]United Institute of Informatics Problems NAS of Belarus,
Filatova str. 7 – 28, Minsk 220026, Belarus, E-mail: l_silver@rambler.ru
[2]Department of Mechanics, University of Zilina, Slovakia
[3]Faculty of Engineering Sciences, Martin-Luther-Universitet Halle-Wittenberg, Germany
[4]Rolls-Royce Centre and Dynamics Group, the University of Sheffield, UK
[5]Composite Nano/Materials Research Laboratory, University of New Orleans, USA

**ABSTRACT**

An adventure at engineering design and modeling is possible with a Virtual Reality Environment (VRE) that uses multiple computer-generated media to let a user experience situations that are temporally and spatially prohibiting. In this paper, an approach to developing some advanced architecture and modeling tools is presented to allow multiple frameworks work together while being shielded from the application program. This architecture is being developed in a framework of workbench interactive tools for next generation nanoparticle-reinforced damping/dynamic systems. Through the use of system, an engineer/programmer can respectively concentrate on tailoring an engineering design concept of novel system and the application software design while using existing databases/software outputs.

## 1. INTRODUCTION

In the past decade, nanoscale measurement techniques and manufacturing technologies are advancing rapidly; however, only few effective techniques for manipulation of nanoparticles are available for engineering applications. Each method has strengths and weaknesses [1] and most of them are generally limited to imaging in two dimensions and hardly used for manipulation of nanoparticles.

Manipulation of nanoscale objects can be achieved using a scanning probe microscope, such as an AFM (Atomic Force Microscope) or an STM (Scanning Tunneling Microscope). Guthold et al. [2] performed rolling and sliding of carbon nanotubes using an AFM. All of these research efforts used a graphical display in addition to the haptic display. The limitation of this approach is that while manipulating the specimen, the graphical display is static and requires additional scans to see the result of the manipulation.

Eigler and Scheweizer [3] used an STM to position individual atoms on a surface, but these tools do not yet offer very precise control of nanoparticles (e.g., the ability to grasp and release). Laser beams can also be used to trap and manipulate small particles. A laser apparatus, called OT (optical tweezers) [4], provides the user with a non-contact method for manipulating objects that can be applied to nanoscale viruses, bacteria, living cells and particles. Although 3D imaging has been reported in [5]. All these techniques require complex driving electronics under sophisticated computer control.

Various techniques in virtual reality environment can be used to enhance human guided computer system. Among these techniques, haptic display is important. One of the most challenging problems in nanoscale manipulation is that real-time image sensing of nanoscale artifacts are difficult. The operation of nanoscale manipulators can be done through tele-operation or automatic manipulation. Hollis et al. [6] demonstrated that atomic-scale landscapes could be explored and felt with the hand in real time by interfacing a haptic feedback device with an STM.

Over the past decade, computer simulation and modeling have become powerful tools for understanding the structure and properties of a broad range of materials [7]. Traditional 3D orthographic displays [8] do not always provide the user with a natural or intuitive way to interact with nanosystems, because of the complexity, novelty, and a scale of systems relevant to nanotechnology. We used some of the advances in modeling approaches to address the problems.

In developing these constraints and rules, one must have some understanding of nanoscale dynamics, but this is still an emerging field of research. This is why tele-operation is gaining favor for certain user interfaces. Tele-operation needs to virtually present the nanoscale environment to the operator such that the operator can be aware of the target objects and the surroundings.

With developing of powerful computers, there have been extensive efforts on virtual prototyping, especially for large-scale systems in areas of transportation (automobile, marine, aerospace and railway). Some software packages are also available for multi-scale modeling, such as ProEng, ANSYS, MatLab, ABAQUS, UGII or AutoCAD. The constructed CAD models can be converted into polygonal models and then are imported in a VRE where various physical behaviors and computation models could be added to the visual mock-ups for real time interaction, dynamic visualization and analysis. This approach has been and is still being widely used at the moment, but many researchers [1-8] have addressed the issue of CAD models, computational tools and VRE integration. It is worth noting some aspects/problems in CAD/CAM of nanoscale systems:

1. Hardware issues. Current CAD/CAM systems are mainly based on complex surface and solid modeling schemes with a big graphical resolution for precise design. Some VRE systems are, therefore, a result of huge databases and require performing computing-extensive operations. Some other obstacles are due to hardware limitations of today's VR systems, such as the accuracy and stability of 3D input and output devices.

2. Modeling issues. Another important reason is the absence of a suitable model representation due to lack of knowledge that would support efficient solid modeling in a VR environment. In addition, traditional CAD systems are often parametric-based modeling systems with exact dimensions and these dimensions typically require precise defining across the length scales at the conceptual design stage.

3. Topological relationships and design conditions could be lost during the translation process of models from CAD systems to VRE systems. Thus to modify a graphical or computational model in a virtual environment, one must go back to computer code systems for making necessary changes and then re-import the modified models into VRE systems for further operations.

Advanced development in computer area should also foster great improvement in VRE development. A suitable VRE-based CAD/CAM workbench tools would satisfy the requirements on both sides of CAD modeling and VR in the following aspects:

*Visualization.* Fast creation and reproducing of required number of convincing and realistic damping elements at nanoscale and then integrating them at macrolevel at high resolutions for quality visualization,

*Interaction.* Intuitive and precise interaction across the length scales,

*Damping behavior simulation.* Simulation of realistic behaviors of virtual objects,

*Modeling.* The support of various computational modeling methods and realistic simulation of design processes.

Our joint group is exploring the use of spatially immersive virtual reality systems and modern IT technologies (e.g., C# Visual Studio Microsoft.Net, Rational Rose and JAVA2 applets) for interactive modeling and visualization of nanoparticle-based systems/materials. A computational scheme and software, which utilizes neural networks and/or Microsoft.Net technique, was developed to predict properties of nanoparticle-reinforced materials and optimization and control of nano-devices.

The principal objective of the paper is to demonstrate an application of modern software engineering tools for modeling virtual reality and molecular dynamics of novel nanocomposites. The main technical components of presented system are engineering workbench for modeling and 3D images of novel nanoparticle-reinforced composites. Recent advances in computer modelling and simulations of the mechanics of materials at the nano and micro scales are reviewed. The Multiscale Modeling of Materials (MMM) approach is shown to rely on systematic reduction of the degrees of freedom at natural length scales. Connections between such scales are currently achieved by either a parameterization or a coarse graining procedure. Parameters that describe the system at a lower length scale are obtained from computer simulations, often verified experimentally, and passed on to upper scales. Alternatively, lower length scale descriptions can be coarse-grained through a "zoom-out" process.

## 2. MODEL AND METHOD OF 2D/3D SIMULATION

The Lagrange approach is used. 2D problems under plane strain and plane stress conditions were considered. The usual set of equations of dynamics of deformable solids for the case of two dimensional plane elastic-plastic flow with von Mises yield criterion was used [9, 10]. The basic equations for strain rates are as follows:

$$\dot{\varepsilon}_{xx} = \frac{\partial v_x}{\partial x}, \quad \dot{\varepsilon}_{yy} = \frac{\partial v_y}{\partial y}, \quad \dot{\varepsilon}_{xy} = \frac{1}{2}\left(\frac{\partial v_y}{\partial x} + \frac{\partial v_x}{\partial y}\right),$$

$$\dot{\omega}_z = -\dot{\omega}_{xy} = \frac{1}{2}\left(\frac{\partial v_y}{\partial x} - \frac{\partial v_x}{\partial y}\right), \quad \dot{\varepsilon}_{zz} = \frac{\dot{h}}{h}$$

- for plane stress state, $\varepsilon_{zz} = 0$ - for plane strain state.

Equations of motion in plane x-y coordinates can be written as follows

$$\rho \frac{\partial v_x}{\partial t} = \frac{\partial \sigma_{xx}}{\partial x} + \frac{\partial \sigma_{xy}}{\partial y}, \quad \rho \frac{\partial v_y}{\partial t} = \frac{\partial \sigma_{xy}}{\partial x} + \frac{\partial \sigma_{yy}}{\partial y}$$

Equation of continuity is $\frac{\dot{V}}{V} = \dot{\varepsilon}_{xx} + \dot{\varepsilon}_{yy} + \dot{\varepsilon}_{zz}$.

Equations of state is
$$\sigma_{ij} = -P\delta_{ij} + s_{ij}, \quad \dot{\varepsilon}_{ij} = \dot{\varepsilon}_{ij}^e + \dot{\varepsilon}_{ij}^p, \quad \dot{\varepsilon}_{ij}^p = \dot{\lambda} s_{ij},$$

Nomenclature: $x, y$ are space coordinates, $v_x$ is velocity in x direction, $v_y$ is velocity in y direction, $\sigma_{xx}, \sigma_{yy}, \sigma_{xy}, \sigma_{zz}$ are stress tensor components, $s_{xx}, s_{yy}, s_{xy}, s_{zz}$ are stress deviators, $\varepsilon_{xx}, \varepsilon_{yy}, \varepsilon_{xy}, \varepsilon_{zz}$ are strain tensor components.

For numerical solving this set of equations a computer program based on the finite-difference scheme known as Wilkins method was used [9, 10]. An algorithm of splitting of grid nodes was applied to model fracture [11, 12]. Stresses of increased accuracy and smooth over the whole domain can be obtained from the nodal displacements and traction boundary conditions using T-functions as interpolators in connection with the Moving Least Squares (MLS) method [8]. We used full second order polynomials and thus for stress evaluation in each point of interest we need displacements in at least 5 nodal points for 2D problems and in 9 nodal points in 3D problem. Near the boundaries, additional conditions are obtained from boundary traction components in the closest boundary point, which increases the accuracy of computed stress field and decreases the number of necessary displacement nodal points in the domain of interest (DOI) for the interpolation. The displacements in internal nodal points, which are necessary for computation of the stress field inside the domain, are obtained from boundary tractions and displacements using the known boundary integral representation [11].

A map of such mesovolume is submitted in fig. 1. There are more than 120 grains with average size about 3 microns in this volume and it can be called representative volume. In calculations for fragments of different colour (grain) the yield strength was different (up to 30%). In uniaxial tension of such mesovolume the system of localized deformation bands take place with inclination of them about 45° to the axis of tension. In fig. 4 right side, the greater values of intensity of plastic deformations

$$\varepsilon_i^{pl} = \sqrt{\frac{2}{3} \left( \varepsilon_{xx}^{pl\,2} + \varepsilon_{yy}^{pl\,2} + 2\varepsilon_{xx}^{pl\,2} + \varepsilon_{zz}^{pl\,2} \right)}$$

correspond to greater intensity of colouring. The deformation in bands, covering on width about 0.5-0.8 microns, ranges up to 30% at integral deformation of a sample 0.7%.

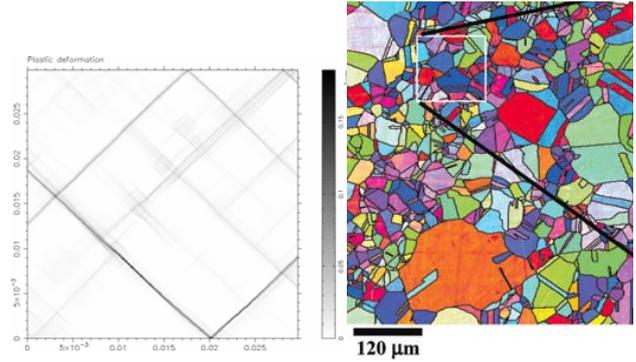

Fig. 1. Map of representative mesovolume and distribution of plastic strains

### 3. ENGINEERING WORKBENCH TOOLS

Overall, the Class Diagram for an engineering workbench is the most important for further synthesis of computer code and programming. It is illustrated in Figs. 1-2. When utilizing an Object-Oriented design process, it is a common practice to draw a sequence diagram for each use case. In doing so, objects and the messages that are sent between objects are defined by user's options 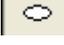 inside of the material/system and actors  outside.

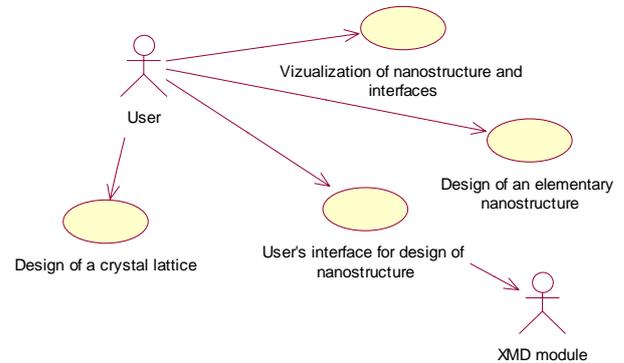

Fig. 2. "Use case" diagram for CAD/CAM of nanostructures

This is useful for process-intensive applications, for the application described in this paper; this stage was skipped going directly to the class diagram. Class diagram-to-use case conformance was checked throughout the design process to verify that the classes were sufficient to implement the use cases. Designed to be modular and extensible, the VRE engineering workbench for damping can be described by two important concepts, functionality and generality. The architecture functionally divides itself into Model, Input, Output and Manager.

Other diagrams can be developed. Component diagram is shown in fig. 2, where  is databases/modules, 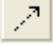 is relations between modules.

The module for design of en elementary nanostructure is based on user's data to be chosen from database or inputted by hands: type of lattice and basis. The module for design of crystal lattice of damping material interprets the data obtained from the above module and then designs some structure of material matrix. The module for visualization treats the data of the former module and graphically represents structure of the material. There is also the module to provide a data transfer from external software such as Matlab, ANSYS etc. so as to grab necessary engineering data into complex VRE workbench tools.

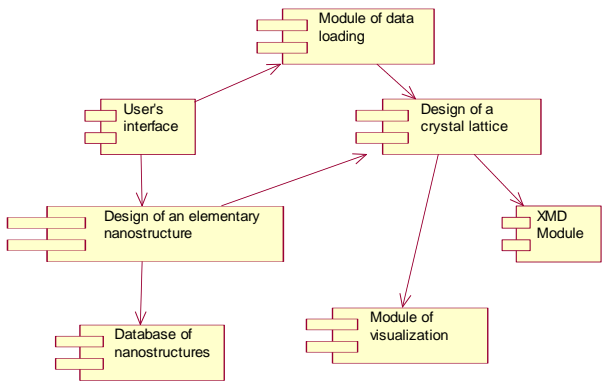

Fig. 2. "Component" diagram

Based on the Class diagram in Rational Rose environment a user can develop computer code of the class on selected programming language. In the case of Microsoft Visual C++ a user can assess entire hierarchy of MFC library classes by using visualized tools that is called - Model Assistant. At some final stages of workbench development, his modules are implemented in Microsoft Visual C++.

The positions of all atoms in the atomic lattices can be further calculated and stored in a data file at two regimes: simulation and calculation.

Regime of simulation is used for visual design purposes in order to find some optimal orientation, distribution and location of particles in a matrix. The design concept is then being used for computational analysis and calculations of damping/dynamic properties of material/system based on affiliated prediction models integrated into database. Virtual database of visual structures and computational tools will introduce outstanding possibilities to researchers in this field.

Resulting picture (fig. 3-5) of nano simulation shows respectively, for example, crystallise lattice of aluminium matrix in fig. 3, diamond nanoparticles of chosen shape in fig. 4 (pyramid-like, sphere or fullerene-like shapes) and nanocomposites represented as aluminium matrix with two introduced pyramid-like diamond nanoparticles (green-atoms in fig. 5). The positions of all atoms in the atomic lattices are also calculated and stored in a data file. The data are used for further calculations of mechanical properties by FEM-based techniques. Virtual database of various nanostructures and their computer-based calculations will introduce outstanding possibilities to researchers in this field. The user can easily attach his hand to an atom or molecule and manoeuvre it in 3D space.

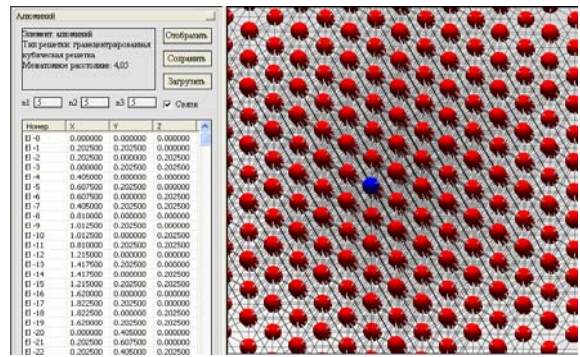

Fig. 3. Atomic lattice of metal matrix

In order to be able to compute large structures containing several nanotubes, the number of degrees of freedom of the overall model has to be reduced. Shell elements could be used for that purpose, see e.g. [14]. Fig. 6 and 7 show the differences between these finite element models.

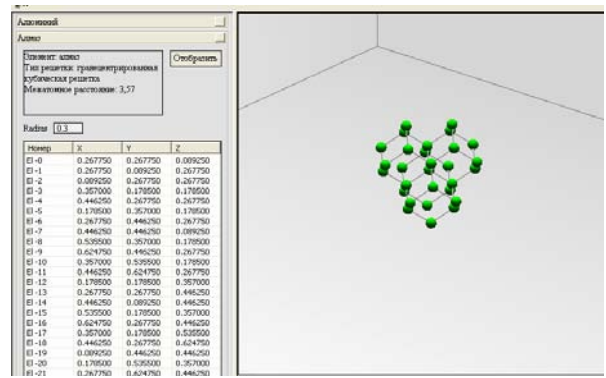

Fig. 4. Atomic lattice of carbon-based nanoparticle

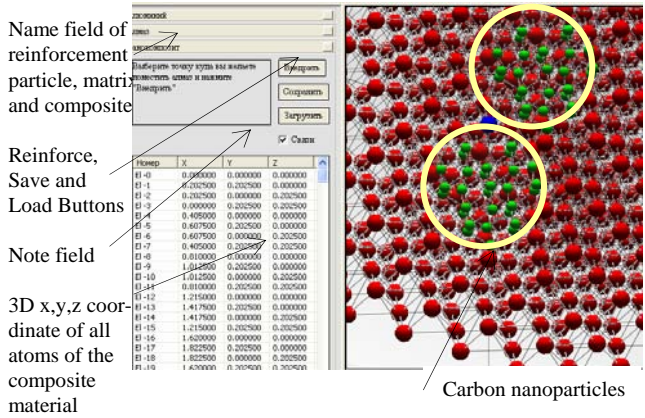

Fig. 5. Reinforced nanocomposite: metal matrix – carbon-based nanoparticles

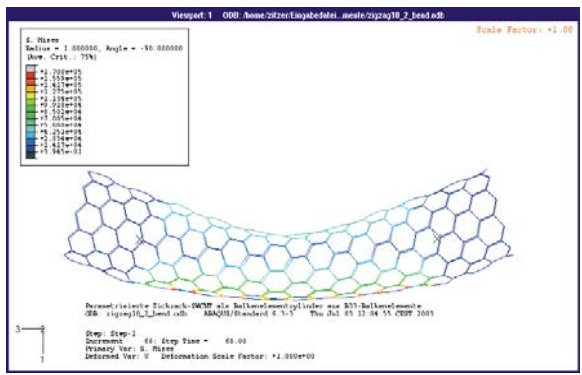

Fig. 6. Beam element model

For a realistic simulation of the mechanical behaviour of nanostructures, the nonlinear intramolecular interactions between neighboring atoms have to be taken into account. In order to reduce computational costs, it is necessary to develop suited techniques, so that shell elements can be applied.

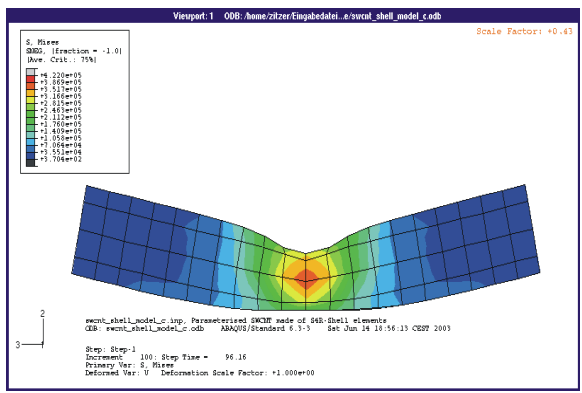

Fig. 7. Shell element model

## 4. CONCLUSION

Computational simulation and modeling tools called as a Virtual Reality Environment (VRE) can help to understand many physical effects and predict the behavior of materials and machine components via computer-generated media. In the present paper, multiscale computational approaches to modeling of nanoparticle-reinforced composite materials and virtual reality engineering tools have been used to describe/model an intuitive interface of some CNT-reinforced materials to enable efficient design and synthesis of next generation materials and nanoscale devices. The underlying mechanics of material has been partially simulated by the use of Lagrange mechanics and programmed accordingly. Hap tic feedback is used to constrain the steering motion within the physical capability of the potential field. In the virtual working environment, the user can naturally grab and steer a nanoparticle, matrix and composite because the information flow between the user and the VRE is bidirectional and the user can influence the environment.

Architecture of VRE is presented which is designed to let multiple computer frameworks work together while being shielded from the application program. The software development of a VRE requires orchestrating multiple peripherals and computers in a synchronized way in real time. Results of the research work will provide a computational platform for the development of nanoparticle-reinforced materials that are lightweight, vibration and shock resistant. The outcome of the research work is expected to have wide-ranging technical benefits with direct relevance to industry in areas of transportation and civil infrastructure development; however, the goal is the next generation of computational and modeling tools for composite nanomaterials.

Multiscale computer modeling of nanocomposites are effective for both theoretical study and educational purposes. Computer simulation may couple various simulation techniques and bridge the length scales together. In the frameworks of classic elastic-plastic model by taking into account the heterogeneous inner structure of a material in explicit form and stress concentrators of various nature, it is possible to simulate numerically regions of localized plastic strain of meso scale that are observed in experiments. Heterogeneity of stressed state is typical for deformation of mesovolumes of a structurally inhomogeneous material. This is due to stress concentrators of various nature and scale (interfaces of fragments of internal structure, feature of the shape etc.). In these conditions, the plastic deformation proceeds heterogeneously. They arise in the region of stress concentration and in the least strength elements of structure. Then bands of localized shear are formed where the plastic deformations much exceed average deformations.

## 5. ACKNOWLEDGEMENTS

Support of the research work by the Royal Society in the United Kingdom and the WELCH scholarship administered through the Amer. Vac. Soc. / Int.-l Union for Vac. Sc., Tech. and Applications in the U.S.A. and Europe is gratefully acknowledged. Dr. Bochkareva is currently continuing her research work under the EU INTAS 2005-2007 postdoctoral fellowship Ref. Nr 04-83-3067. Further research work is also being supported by Marie-Curie Fellowship Ref. # 021298-Multiscale Damping at the Rolls-Royce Center in Damping Technology, the University of Sheffield in the United Kingdom. It should be noted however that the views expressed in this paper are those of the authors and not necessarily those of any institutions.